\magnification=1200
\vbadness=10000

\hfuzz=10pt \overfullrule=0pt
\baselineskip=12pt
\parindent 20pt \parskip 6pt
\def\Tr{\mathop{\rm Tr}\nolimits}
\def\mapright#1{\smash{\mathop{\longrightarrow}\limits^{#1}}}

\def\su#1{{\rm SU}(#1)}
\def\so#1{{\rm SO}(#1)}

\def\u#1{{\rm U}(#1)}
\def\p#1{{\pi_{#1}}} 
\def\Z{{\bf Z}}
\def\R{{\bf R}}
\def\M{{\cal M}}
\def\c{c_\chi}
\def\s{s_\chi}
\def\xh{\hat x}
\def\yh{\hat y}
\def\zh{\hat z}

\hfill\vbox{\hbox{{hep-ph/9701351}}
\hbox{{DAMTP-R-97/04}}
\hbox{{NSF-ITP/97-013}}}
\vskip .3in

\centerline{{\bf THE COLLAPSE OF EXOTIC TEXTURES}}
\vskip .2in

\centerline {Andrew Sornborger,$^{(1)}$ Sean M. Carroll,$^{(2)}$ and 
Ted Pyne$^{(3)}$}
\vskip .3cm
\centerline{\it $^{(1)}$Department of Applied Mathematics and
Theoretical Physics}
\centerline{\it University of Cambridge}
\centerline{\it Silver Street, Cambridge CB3 9EW}
\centerline{\it Great Britain}
\centerline{\it email: {\tt A.T.Sornborger@damtp.cam.ac.uk}}
\vskip .3cm
\centerline{\it $^{(2)}$Institute for Theoretical Physics}
\centerline{\it University of California}
\centerline{\it Santa Barbara, California\quad 93106}
\centerline{\it email: {\tt carroll@itp.ucsb.edu}}
\vskip .3cm
\centerline{\it $^{(3)}$Harvard-Smithsonian Center for Astrophysics}
\centerline{\it Cambridge, Massachusetts\quad 02138}
\centerline{\it email: {\tt pyne@cfa160.harvard.edu}}

\vskip .3in

\centerline{\bf Abstract}
\vskip .1in

The ordering of scalar fields after a phase transition in which a group 
$G$ of global symmetries is spontaneously broken to a subgroup $H$
provides a possible explanation for the origin of structure in the 
universe, as well as leading to observable effects in condensed
matter systems.
The field dynamics can depend in principle on the geometry and 
topology of the vacuum manifold $G/H$; for example, texture configurations
which collapse and unwind will exist if the third homotopy group
$\pi_3(G/H)$ is nontrivial.  We numerically simulate the evolution of 
texture-like configurations in a number of different models, 
in order to determine the extent to which the geometry
and topology of the vacuum manifold influences the field evolution. 
We find that the dynamics is affected by whether or not the theory
supports strings or monopoles [characterized by $\pi_1(G/H)$
and $\pi_2(G/H)$, respectively].  In some of the theories studied,
configurations with initially spherically symmetric energy densities
are unstable to nonspherical collapse; these theories are also found
to nucleate defects during the collapse.
Models that do not support monopoles or strings behave similarly 
to each other, regardless of the specific vacuum manifold.

\vfill
\noindent{January 1997}

\eject

\baselineskip 20pt

\noindent{\bf I. Introduction}

Our understanding of large-scale inhomogeneities in the universe, both
directly through surveys of the distances and velocities of galaxies 
and indirectly through observations of temperature fluctuations in the 
cosmic microwave background, has advanced significantly in recent years.
This growth in empirical knowledge has necessitated greater precision
in the extraction of predictions from theoretical models; claiming that
a theory is consistent with observation requires more care
now than it did a decade ago.  As a result, it has become important
to pay close attention to the differences between various manifestations
of any general scenario.  This paper takes up this task for the texture
model of structure formation, and studies the extent to which the geometry 
and topology of the vacuum manifold in a scalar field theory can exert
a cosmologically significant influence on the dynamics of the fields.
Although our discussion is phrased in terms of cosmological effects,
analogous considerations hold for condensed matter systems with
spontaneously broken global symmetries.

In the texture scenario, we consider a set of scalar fields $\Phi$ 
which transform under a global symmetry group $G$.  If the potential
$V(\Phi)$ is such that this symmetry is spontaneously broken to a
subgroup $H$, the potential will possess a set of degenerate
minima.  This set of field values is known as the vacuum
manifold $\M$, and is isomorphic to the quotient space $G/H$.  Typically,
at high temperatures the expectation value of the field is the same
everywhere; as the universe cools the fields will relax to the 
vacuum manifold, and according to the Kibble mechanism the points to
which the field evolves will be uncorrelated on distances greater than
the Hubble distance at that epoch.  The fields will tend to order
themselves, approaching a constant-field configuration within a causally
connected region, but as the Hubble distance grows the fields will always
be uncorrelated on cosmological scales.  The resulting gradient energy
leads to an approximately scale-free spectrum of energy density
perturbations, which can serve as seeds for large-scale structure.

Depending on the topology of the vacuum manifold, the scalar fields
may be forced to leave $\M$ in order to smooth themselves out.  This can
be seen by considering a configuration which is confined to $\M$, and
is set to some fixed value outside a
certain radius.  This is mathematically equivalent to compactifying
three-dimensional space to a three-sphere, and the resulting field
configuration defines a map $S^3 \rightarrow \M$.  Such maps are
classified topologically by the third homotopy group $\p3(\M)$; if the
configuration corresponds to a nontrivial element of $\p3(\M)$, then it
cannot be smoothly deformed to a constant-field configuration.  In
this case, it is energetically favorable for the configuration to shrink
in size until the energy density reaches that needed to leave
the vacuum manifold.  The field then ``unwinds'' by climbing over the
energy barrier to a topologically trivial configuration.
Turok [1] suggested that this process of collapse and unwinding could
lead to seeds for the formation of large-scale structure.  It was later
realized [2,3] that scalar field gradients could lead to density
perturbations even if $\p3(\M)$ were trivial.

Even though density perturbations will be produced for any theory with
spontaneously broken exact global symmetries, 
it is reasonable to imagine that the detailed 
dynamics of the fields (and hence the specific prediction for
perturbations) will depend on the topology of the vacuum manifold.
Indeed, two different models with nontrivial $\pi_3(\M)$, which
support texture configurations with nonzero winding, may nevertheless
predict different evolutions for the collapsing texture, and such
differences may manifest themselves cosmologically.  At the same time,
it may be possible for configurations which are topologically trivial
to evolve in ways similar to what we would call a texture.

In this paper we explore some of these issues.  We consider
theories of scalar fields transforming under global symmetry
groups; the theories are specified by the group $G$, the
representation of $G$ under which the fields $\Phi$ transform, and the
potential energy $V(\Phi)$.  Li [4] has considered a number of such
theories, and determined what the unbroken symmetry group $H$ will be
in each model.  Bryan, Carroll and Pyne [5] have examined the topology
of the resulting vacuum manifolds $\M=G/H$, and calculated the
homotopy groups $\p1(\M)$, $\p2(\M)$, and $\p3(\M)$.  (Just as
$\p3(\M)$ characterizes textures, $\p1(\M)$ characterizes cosmic 
strings and $\p2(\M)$ characterizes monopoles.)  Here we have chosen
eight different models, listed in Table One, which represent a
variety of different field contents and vacuum manifold topologies.

To understand in full detail the cosmological effects of each of 
these theories, it would be necessary to do a number of full-scale 
numerical simulations of large-scale structure formation or CMB
anisotropy in each
model, such as those which have been performed for the models with
$\M=S^3$ [6,3].  We have adopted a more modest approach, since our goal 
is simply to determine whether the choice of vacuum manifold exerts
an influence over the dynamics.  Our strategy is therefore to set up
comparable field configurations in the various theories, each corresponding 
to a single texture.  We then follow numerically the evolution of the
configurations as they collapse and evaporate away, keeping track of
the distribution of energy into potential, kinetic and gradient
energy, as well as the asymmetry of the collapse as measured by the
quadrupole moments of the total energy density.  (For analogous 
studies of the usual texture models, see [7,8].)  In each model we
performed three simulations:  one in which the energy density was
initially spherically symmetric, one which was initially prolate 
(cigar-shaped), and one which was initially oblate (pancake-shaped).

As detailed below, we find that the models studied fall naturally
into two different categories of behavior.  Within each of the two
classes, the evolution of the energy densities and quadrupole moments
is relatively similar for each type of collapse, while there is a
noticeable difference between the two categories.  In the eight
models we studied, the division corresponds precisely to whether
the theory supports strings or monopoles; {\it i.e.}, to whether or 
not $\p1(\M)$ and $\p2(\M)$ are trivial.  All of the models without these
defects evolved in one way, while those which do support defects
evolved in a distinct fashion.  Upon examination of the simulations,
we discover that in some of the models defects are nucleated in
the process of collapse; however this is not always true, and is
therefore not a sufficient explanation for the division into two
types of collapse.

\noindent{\bf II. Initial Configurations}

The models we consider are described by Lagrangians of the form
$$
  {\cal L} = {1\over 2}\partial_\mu\Phi\cdot\partial^\mu\Phi
  -V(\Phi)\ .\eqno(2.1)
$$
Here, $\Phi$ stands for a collection of $N$ scalar fields which
transform under some representation of a global symmetry group $G$.
The product $\Phi_1\cdot\Phi_2$ is the invariant inner product
appropriate to the representation; for a vector $\Phi^a$ we have
$\Phi_1\cdot\Phi_2 = \sum_a \Phi_1^{a*} \Phi_2^a$ (where an
asterix denotes complex conjugation), while for
a matrix $\Phi^{ab}$ we have $\Phi_1\cdot\Phi_2 = \sum_{ab}
\Phi_1^{ab*} \Phi_2^{ab} = \Tr(\Phi^\dagger_1 \Phi_2)$.  The potential
$V(\Phi)$ consists of terms quadratic and quartic in the fields.

We are interested in constructing, for each theory,
initial field configurations which are wholly in the vacuum manifold 
and which have unit winding number.  We follow the procedure outlined
in [5].  The basis of this procedure is the fact that every field
value in the vacuum manifold may be obtained by starting with a 
fixed vacuum expectation value $\langle\Phi\rangle$ and acting on
it by an element of the symmetry group $G$.  The initial configuration
can therefore be written
$$
   \Phi({\bf x}) = \mu({\bf x})\langle\Phi\rangle\ ,\eqno(2.2)
$$
where $\mu({\bf x})$ is a position-dependent element of $G$.  (The
notation $\mu({\bf x})\langle\Phi\rangle$ indicates the action of $\mu$ 
on the representation carried by $\langle\Phi\rangle$; for vector 
representations it
will be matrix multiplication, while for matrix representations it
will be conjugation.)  By considering transformations $\mu({\bf x})$ 
which go to the identity at spatial infinity, $\mu$ defines a map
$\mu: S^3\rightarrow G$, and hence an element of $\p3(G)$.  Elements
of $\p3(G/H)$ are related to those of $\p3(G)$ by the exact homotopy
sequence
$$
  \matrix{\p3(G)&\mapright{\beta}&
    \p3(G/H)&\mapright{\gamma}& \p2(H)\ .\cr}\eqno(2.3)
$$
Since $\p2(H)=0$ for any $H$, exactness implies that every element
of $\p3(G/H)$ is the image under $\beta$ of some element of 
$\p3(G)$.  To construct a configuration with winding number one in
the vacuum manifold, it is therefore necessary to find a map $\mu$
with an appropriate winding number in $\p3(G)$.

The map $\mu$ can always be written as a composition,
$$
  \mu=\alpha\circ\widetilde\mu\ ,\eqno(2.4)
$$
where $\widetilde\mu$ is a fixed
map from $S^3$ to $\su2$ and $\alpha$ maps $\su2$ to $G$.  If we
choose $\widetilde\mu$ to have winding number one, then $\mu$ will
represent the same winding number as $\alpha$.  For $\widetilde\mu 
:S^3\rightarrow\su2$ we choose
$$
  \widetilde\mu({\bf x}) = \left(\matrix{ \c+i\zh\s &
  (i\xh +\yh)\s \cr (i\xh -\yh)\s & \c-i\zh\s\cr}\right) \ .\eqno(2.5)
$$
Here we have introduced the notation 
$\c = \cos(\chi)$, $\s = \sin(\chi)$, 
$\chi(r)$ a function of $r=\sqrt{ x^2+y^2+z^2}$ with
$\chi (0)=0$ and $\chi (\infty )=\pi$, and $\hat{\bf x} = {\bf x}/r$.
Construction of an appropriate field configuration then comes down
to choosing the map $\alpha :\su2\rightarrow G$.

We begin with a theory in which $\so4$ breaks down to $\so3$.  In
this theory the vacuum manifold is $\so4/\so3 = S^3$; it is
the simplest theory containing textures, and may be thought of as the
standard against which other models should be compared.  The field 
content consists of four real scalars $\Phi^a\ 
(a=1,\dots 4$), transforming under a vector representation of $\so4$.
The action of an $\so4$ matrix $O^a{}_b$ is thus $\Phi^a \mapsto
O^a{}_b\Phi^b$.  The potential is given by
$$
  V(\Phi^a) = \lambda(\Phi^a\Phi^a -v^2)^2\ .\eqno(2.6)
$$
This potential is minimized when the fields attain a vev of the form
$$
  \langle\Phi\rangle = \left(\matrix{v\cr 0\cr 0\cr 0\cr}
  \right)\ ,\eqno(2.7)
$$
breaking $\so4$ to an $\so3$ subgroup, namely the subgroup consisting of
$4\times 4$ matrices with a one in the upper left corner, zeroes
in the rest of the first row and column, and $\so3$ matrices
in the lower right $3\times 3$ subblock.  
The map $\alpha$
must now be a homomorphism from $\su2$ to $\so4$ with winding
number one.  [In fact $\p3(\so4)=\Z\oplus\Z$, so the concept of
winding number is not uniquely defined.  The map we choose is
that which induces the generator of $\p3(\so4/\so3)$.]
If we write a general element of $\su2$
in terms of four real parameters $e_a$ with $\Sigma e_ae_a=1$,
$$
  g=\left(\matrix{e_0+ie_3&e_2+ie_1\cr -e_2+ie_1&e_0-ie_3\cr}
  \right)\ ,\eqno(2.8)
$$ 
then an appropriate map is given by
$$
\alpha (g)= \left(\matrix{ e_0&e_3&e_2&e_1\cr
  -e_3&e_0&-e_1&e_2\cr -e_2&e_1&e_0&-e_3\cr
  -e_1&-e_2&e_3&e_0\cr}\right)\ .\eqno(2.9)
$$ 
We may think of $\alpha$ as replacing complex entries by real
$2\times 2$ matrices:
$$
  a+ib \mapsto \left(\matrix{a&b\cr -b&a\cr}\right)\ .\eqno(2.10)
$$
The composition of $\alpha$ with $\widetilde\mu({\bf x})$
results in a map $\mu({\bf x})$ given by
$$
  \mu({\bf x}) = \left(\matrix{ \c&\zh\s&\yh\s&\xh\s\cr
  -\zh\s&\c&-\xh\s&\yh\s\cr -\yh\s&\xh\s&\c&-\zh\s\cr
  -\xh\s&-\yh\s&\zh\s&\c\cr}\right)\ ,\eqno(2.11)
$$
which leads in turn to a field configuration
$$
  \Phi({\bf x}) = v\left(\matrix{\c\cr -\zh\s\cr -\yh\s\cr
  -\xh\s\cr}\right)\ .\eqno(2.12)
$$

The model in which $\so5$ breaks down to $\so4$ is closely related
to the previous one.  The fields are arranged into a five-vector
$\Phi^a$, with potential as in (2.6).
The vacuum manifold is $\so5/\so4=S^4$,
for which $\p3(S^4)=0$; there are therefore no configurations with
nonzero winding number.  Nevertheless, we can set up configurations
which look just like those in the previous theory and see how they
evolve.  This serves as a test of the importance of the topology
of the vacuum manifold in texture scenarios.

The configurations we
consider are maps from $S^3$ to $S^4$ which lie at a constant distance
from the north pole in $S^4$.  They are given by
$$
  \Phi({\bf x}) = v\left(\matrix{a\c\cr -a\zh\s\cr -a\yh\s\cr
  -a\xh\s\cr \sqrt{1-a^2}\cr}\right)\ .\eqno(2.13)
$$
The parameter $a$ determines the distance from the north pole; $a=1$
corresponds to the equator of $S^4$, while $a=0$ puts the entire
field configuration at the pole (a single point).

Another closely related theory features $\so5$ breaking down to
$\so3$.  This model provides an interesting comparison with
$\so4/\so3$, since in this case $\p3(G/H)$ is the finite group
$\Z_2$.  There are ten real scalar fields, arranged into two
five-vectors $\Phi_1^a$ and $\Phi_2^a$, with potential
$$
  V\left( \Phi_1^a,\Phi_2^b \right)= \lambda_1 \left(\Phi_1^a
  \Phi_1^a -v_1^2\right)^2+
  \lambda_2 \left(\Phi_2^b\Phi_2^b -v_2^2\right)^2+\eta\left(
  \Phi_1^a\Phi_2^a\right)^2\ .\eqno(2.14)
$$
The fields acquire vevs of the form
$$
  \langle\Phi_1\rangle =\pmatrix{v_1\cr 0\cr 0\cr 0\cr 0\cr}\ ,\ \ 
  \langle\Phi_2\rangle =\pmatrix{0\cr v_2\cr 0\cr 0\cr 0\cr}\ .
  \eqno(2.15)
$$
We act on these fields by a $5\times 5$ matrix with (2.11) as the
upper left $4\times 4$ subblock, a one in the lower right corner,
and zeroes elsewhere.  (Note that a single $\so5$ matrix acts 
simultaneously on the two vectors.)  The resulting configuration is
$$
  \Phi_1 = v_1 \pmatrix{\c\cr -\zh\s\cr -\yh\s\cr
  -\xh\s\cr 0\cr}\ ,\ \ 
  \Phi_2 = v_2 \pmatrix{\zh\s\cr \c\cr \xh\s\cr -\yh\s\cr 0\cr}\ .
  \eqno(2.16)
$$

We next consider a theory where $\su3$ is completely broken
($H=0$).  (The appearance of textures in a  particle physics model with 
this symmetry-breaking pattern was examined in [9].)
The lower homotopy groups of the resulting vacuum manifold
are the same as those of $\so4/\so3$, although the number of fields
and the dimensionality of the vacuum manifold are both greater.  The 
fields consist of two complex three-vectors $\Phi_1^a$ and $\Phi_2^a$, 
for a total of twelve real degrees of freedom. The potential is given by
$$
  V\left( \Phi_1^a,\Phi_2^b \right)= \lambda_1 \left(\Phi_1^{\dagger}
  \Phi_1 -v_1^2\right)^2+
  \lambda_2 \left(\Phi_2^{\dagger}\Phi_2 -v_2^2\right)^2+\eta\left(
  \Phi_1^{\dagger}\Phi_2+\Phi_2^{\dagger}\Phi_1\right)^2\ .\eqno(2.17)
$$
The fields attain vacuum expectation values of
$$
\langle\Phi_1\rangle =\pmatrix{v_1\cr 0\cr 0\cr}\ \ \ 
\langle\Phi_2\rangle =\pmatrix{0\cr v_2\cr 0\cr}\ .\eqno(2.18)
$$
In this case the map $\alpha :\su2\rightarrow\su3$ is simply inclusion
of a $2\times 2$ matrix into the upper left corner of a $3\times 3$
matrix.  Performing the transformation $\mu=\alpha\circ\widetilde\mu$
on the two vectors then yields
$$
  \Phi_1 = v_1 \pmatrix{\c+i\zh\s \cr (i\xh -\yh)\s \cr 0\cr}\ ,\ \ 
  \Phi_2 = v_2 \pmatrix{(i\xh +\yh)\s \cr \c-i\zh\s\cr 0\cr}\ .
  \eqno(2.19)
$$

The theory in which $\so3$ breaks down to $\so2$ is once again
similar to $\so4$ breaking to $\so3$.  The resulting textures are
known as ``Hopf textures'' [10].  The fields comprise a 
real three-vector $\Phi^a$ with potential of the form (2.6), which
leads to a vev
$$
  \langle\Phi\rangle = \left(\matrix{v\cr 0\cr 0\cr}
  \right)\ .\eqno(2.20)
$$
The vacuum manifold is $\so3/\so2 = S^2$, for which both $\p2$
and $\p3$ are $\Z$.  The theory therefore contains monopoles as
well as textures.
Unlike the two cases above the map $\alpha$ of (2.9)
will clearly not suffice. Instead with $g$ as in (2.8), we set
$$
\alpha(g)=\left(\matrix{ e_0^2+e_1^2-e_2^2-e_3^2&
2\left( e_1e_2+e_0e_3\right)&2\left( e_1e_3-e_0e_2\right)\cr
  2\left( e_1e_2-e_0e_3\right)&e_0^2-e_1^2+e_2^2-e_3^2
&2\left( e_2e_3+e_0e_1\right)
\cr 2\left( e_1e_3+e_0e_2\right)&2\left( e_2e_3-e_0e_1\right)&
e_0^2-e_1^2-e_2^2+e_3^2\cr}\right)\ .\eqno(2.21)
$$  
This map is the standard double cover of $\so3$ by $\su2$.  (The
winding number is nevertheless one; there are no single covers.)
Its composition with $\widetilde\mu({\bf x})$ from (2.5) results in
$$
  \mu = \left(\matrix{
  \c^2+(2\xh^2-1)\s^2 &2(\xh\yh\s^2-\zh\c\s) &2(\xh\zh\s^2+\yh\c\s)\cr
  2(\xh\yh\s^2+\zh\c\s) &\c^2+(2\yh^2-1)\s^2 &2(\yh\zh\s^2-\xh\c\s)\cr
  2(\xh\zh\s^2-\yh\c\s) &2(\yh\zh\s^2+\xh\c\s) &\c^2+(2\zh^2-1)\s^2\cr}
  \right)\ .\eqno(2.22)
$$
The resulting field configuration is straightforward:
$$
  \Phi({\bf x}) = v\left(\matrix{\c^2 + (2\xh^2-1)\s^2 \cr
  2(\xh\yh\s^2+\zh\c\s)\cr 2(\xh\zh\s^2-\yh\c\s)\cr}
  \right)\ .\eqno(2.23)
$$

Adding an additional three-vector to the above theory yields a model
in which $\so3$ is completely broken.  The vacuum manifold is
therefore $\so3$ itself, which supports $\Z_2$ strings as well as
textures.  The potential is of the
form (2.14), leading to vevs of the form
$$
  \langle\Phi_1\rangle =\pmatrix{v_1\cr 0\cr 0\cr}\ ,\ \ 
  \langle\Phi_2\rangle =\pmatrix{0\cr v_2\cr 0\cr}\ .
  \eqno(2.24)
$$
Acting on these fields with (2.22) gives the following field 
configuration:
$$
  \Phi_1({\bf x}) = v_1\left(\matrix{\c^2 + (2\xh^2-1)\s^2 \cr
  2(\xh\yh\s^2+\zh\c\s)\cr 2(\xh\zh\s^2-\yh\c\s)\cr}
  \right)\ ,\quad
  \Phi_2({\bf x}) = v_2\left(\matrix{2(\xh\yh\s^2-\zh\c\s)\cr
  \c^2 + (2\yh^2-1)\s^2 \cr 2(\yh\zh\s^2+\xh\c\s)\cr}
  \right)\ .\eqno(2.25)
$$

We turn now to a model in which an $\so4$ symmetry is broken
to $\u2$.  This is accomplished by a set of six scalars which
transform in the antisymmetric tensor representation of $\so4$,
$\Phi^{ab}=-\Phi^{ba}$.  The action of an $\so4$ transformation
$O$ on the fields is given by $\Phi\mapsto O^T\Phi O$.  The
quartic potential invariant under this symmetry is given by
$$
  V(\Phi^{ab})=-{1\over 2}m^2\Tr(\Phi^{\rm T}\Phi) +{1\over 4} 
  \lambda_1 [\Tr(\Phi^2)]^2 +{1\over 4}\lambda_2\Tr(\Phi^4)\ ,\eqno(2.26)
$$
where $\Phi^n$ refers to matrix multiplication.
The fields acquire a vev of the form
$$
  \langle\Phi\rangle = v\left(\matrix{ 0&1&0&0\cr -1&0&0&0\cr
  0&0&0&1\cr 0&0&-1&0\cr}\right)\ .\eqno(2.27)
$$
In this case we cannot use (2.11) to act on the vev, since this
transformation is actually in the unbroken subgroup $H=\u2$,
and hence leaves (2.27) invariant.  (The generator
of $\p3(\so4/\u2)$ and the generator of $\p3(\so4/\so3)$ are
inherited from the two different generators of 
$\p3(\so4)=\Z\oplus\Z$.) Instead we take the $\so3$
transformation (2.22) and include it in $\so4$ by putting a one
in the lower right corner and zeroes elsewhere.  Acting the result
on (2.27) yields a field configuration
$$
  \Phi = v\left(\matrix{ 0&\c^2 + (2\zh^2 -1)\s^2 &
  -2(\xh\c\s + \yh\zh\s^2) & 2(\xh\zh\s^2 - \yh\c\s) \cr
  -\c^2 - (2\zh^2 -1)\s^2 & 0 & 2(\xh\zh\s^2 - \yh\c\s) &
  2(\yh\zh\s^2 + \xh\c\s) \cr
  2(\xh\c\s + \yh\zh\s^2) & -2(\xh\zh\s^2 - \yh\c\s) &
  0 & \c^2 + (2\zh^2 -1)\s^2 \cr
  -2(\xh\zh\s^2 - \yh\c\s) & -2(\yh\zh\s^2 + \xh\c\s) &
  -\c^2 - (2\zh^2 -1)\s^2 & 0\cr}\right)\ .\eqno(2.28)
$$

Our final model features $\so5$ breaking down to 
$\so3\times\so2\times\Z_2$.  There are fourteen scalar fields
transforming in the symmetric and traceless tensor representation
of $\so5$: $\Phi^{ab}=\Phi^{ba}$, $\Tr\Phi=0$.  The potential is
$$
  V(\Phi^{ab})=-{1\over 2}m^2\Tr(\Phi^2) +{1\over 4} 
  \lambda_1 [\Tr(\Phi^2)]^2 +{1\over 4}\lambda_2\Tr(\Phi^4)\ ,\eqno(2.29)
$$
leading to a vev of the form
$$
  \langle\Phi\rangle = v\left(\matrix{ 2&0&0&0&0\cr 0&2&0&0&0\cr
  0&0&2&0&0\cr 0&0&0&-3&0\cr 0&0&0&0&-3\cr}\right)\ .\eqno(2.30)
$$
The map $\mu ({\bf x})$
we have already found in (2.11). Its action on the vev yields the field 
configuration $\Phi = \mu^{\rm T}\langle\Phi\rangle\mu$, 
which in components is
$$
  \Phi = v\left(\matrix{
  2-5\xh^2\s^2 & -5\xh\yh\s^2 & 5\xh \zh\s^2 & 5\xh\c\s & 0 \cr
  -5\xh\yh\s^2 & 2-5\yh^2\s^2 & 5\yh \zh\s^2 & 5\yh\c\s & 0 \cr
  5\xh\zh\s^2 & 5\yh\zh\s^2 & 2-5\zh^2\s^2 & -5\zh\c\s & 0 \cr
  5\xh\c\s & 5\yh\c\s & -5\zh\c\s & 2-5\c^2 & 0 \cr
  0 & 0 & 0 & 0 & -3\cr}\right)\ .\eqno(2.31)
$$

To compare the different theories, we would like to compare the
evolution of configurations which are in some sense ``the same,''
despite being in different models.  For example, it would be reasonable
to compare the collapse of initially spherically symmetric textures.
In all of the above field configurations, the initial energy density is
spherically symmetric if we choose the radial function
$$
  \chi(r) = 2\arctan(r)\ .\eqno(2.32)
$$
Indeed, with this choice the energy density profiles are identical
(up to an overall normalization); they are given by
$$
  \rho(x,y,z) \propto {1\over{(1+r^2)^2}}\ .\eqno(2.33)
$$
In this sense, the initial configurations we consider are directly
comparable.

However, there is a more restrictive definition of spherical symmetry:  
that a configuration which is rotated about the origin in space can 
always be brought to its original form by a global symmetry transformation.
Not all of the theories we consider allow configurations which
are symmetric in this strict sense.  It is straightforward to check,
by considering infinitesimal spatial rotations and transformations in
the symmetry group $G$, that all of the textures in the models without
defects [$\so4/\so3$, $\so5/\so4$, $\so5/\so3$, and $\su3/0$] are
spherically symmetric, and in addition the configuration in 
$\so5/[\so3\times\so2\times\Z_2]$ is as well.  In the remaining theories
it is impossible to set up a truly spherically symmetric configuration.
For the theories based on $\so3/\so2 = S^2$ and $\so4/\u2 = \R P^2$, 
the vacuum manifolds are two-dimensional (and the map from space to
the vacuum manifold is onto); therefore, the preimage of
any given point in the vacuum manifold will be a one-dimensional
region of space, which in general will not be taken into 
another such one-dimensional preimage under
a spatial rotation.  In the $\so3/0 = \R P^3$ theory, meanwhile,
the spatial rotation has precisely two fixed points (the origin and 
the point at infinity), while the group action has none; therefore, a
rotation cannot be undone by a group transformation.

As we shall see, this lack of true spherical symmetry will manifest
itself in our simulations.  This can be traced to the following effect:
if a configuration is not truly symmetric but has a symmetric energy
density, deformations of the configuration along different coordinate
axes will not yield configurations of equal energy.  Hence, 
there will be an instability to non-spherical collapse.  We will see
this explicitly in our simulations of textures with initially
spherically symmetric energy densities.  Furthermore, we also perform 
simulations of oblate and prolate textures, obtained from those with
symmetric energy densities by deforming along one axis.  For the
truly symmetric initial configurations, the axis chosen is of no
consequence, and the energy density of the deformed texture is simply
the deformed energy density of the symmetric texture.
The configurations which are not truly symmetric, however, have a
preferred direction [the $x$-axis for (2.23), and the $z$-axis for
(2.25) and (2.28), with our conventions].  A deformation orthogonal
to the preferred direction results in a triaxial energy density with
no rotational symmetry, while a deformation along the preferred direction
yields an axially symmetric energy density.  In $\so3/0$ this axially
symmetric pattern is very similar to that of the models which allow
spherical symmetry, while in $\so3/\so2$ and $\so4/\u2$ the resulting
energy density is toroidal.   In Figure One we have plotted 
the energy densities in two-dimensional slices through two representative
deformed configurations, $\so4/\so3$ and $\so3/\so2$.

The energy density profiles of the oblate and prolate configurations 
are therefore not identical in different theories,
and consequently the comparison between the models is not as direct
as in the initially symmetric case.  We will see below, however, that
the difference is not crucial.  For our simulations of deformed textures,
we have always performed the deformation along the preferred axis, so
that the comparison is always between configurations with axially
symmetric energy densities.  

\noindent{\bf III. Numerical Implementation}

We integrate the field equations of motion using a standard staggered
leapfrog algorithm, using a method that is second-order in time but
which uses fourth-order spatial differences (as in [8], except we have
added fourth-order spatial differences for better resolution). The
field equations for the above Lagrangian with a general potential
$V(\Phi )$ are
$$
  {\partial^2\over{\partial t^2}} \Phi - \nabla^2 \Phi +
  {\partial V\over{\partial \Phi}} = 0\ .\eqno(3.1)$$
Here, the field $\Phi$ is vector or tensor valued, depending on which 
theory we are integrating. In the staggered leapfrog method, the momenta 
of the field $\pi = \dot\Phi$ are defined on half-integer
timesteps. Using finite differences as an approximation to the 
differential operators, the expression for the field at
discrete timestep $n + 1$ and Cartesian grid location $x = i \Delta
x$, $y = j \Delta x$ and $z = k \Delta x$ is
$$
  \Phi ^{n + 1}_{ijk} = \Delta t\ \pi^{n + 1/2}_{ijk} + \Phi^n_{ijk}
  \eqno(3.2)
$$
and from the field equations, combined with the momentum definition,
we obtain the expression
$$\eqalign{
  \pi^{n + 1/2}_{ijk} = \pi^{n-1/2}_{ijk} 
  - {\Delta t\over{\Delta x^2}} &
  (\Phi^n_{i+2,jk} + \Phi^n_{i,j+2,k} + \Phi^n_{ij,k+2} 
  - 16 \Phi^n_{i+1,jk} - 16 \Phi^n_{i,j+1,k} \cr
  & - 16 \Phi^n_{ij,k+1}
  + 90 \Phi^n_{ijk} - 16 \Phi^n_{i-1,jk} - 16 \Phi^n_{i,j-1,k} \cr
  & - 16 \Phi^n_{ij,k-1}
  + \Phi^n_{i-2,jk} + \Phi^n_{i,j-2,k} + \Phi^n_{ij,k-2})
  - \Delta t{\partial V\over{\partial\Phi}} }\eqno(3.3)
$$
for $\pi$ at timestep $n + 1/2$.

We use Neumann boundary conditions. In our simulations, once the
initial texture configuration collapses, massless and massive
radiation hits the wall, then some of it is reflected and influences
the subsequent evolution of the field inside the box. In our
production simulations, we take the radius of the texture to be $10$
gridzones, while the simulation volume is $64\times 64\times 64$
gridzones. With this geometry, reflected radiation does not influence
the texture configuration until after it has collapsed and the decay
products (either defects or radiation) have had time to propagate away
from the texture core to a distance of order the initial size.

The initial texture configurations are discussed above for the various
theories which we have simulated.  For the purposes of the numerical
simulation it is convenient to have an initial condition with
zero energy at the boundaries of the simulation volume; we therefore
modify the radial function $\chi(r)$ given in (3.32) so that the
fields are initially constant outside a fixed radius:
$$
  \chi(r) = \cases{2
  \left[1 - e^{\beta (r - R)}\right] \arctan(r)&  for $r \leq R$ ,\cr
  \pi & for $r > R$\ , \cr}\eqno(3.4)
$$
with $\beta$ and $R$ set such that $\chi$ goes smoothly to $\pi$ at
the finite radius $R$.  We also examined
textures in a theory where $\su3$ was spontaneously broken to $\so3$;
in this model, however, no natural choice of $\chi(r)$ led to a
symmetric energy density.  We have therefore not included this model
in our comparisons.

There are two conditions on the timestep for a global symmetry
breaking theory. One is the usual Courant condition $\sqrt{2} \Delta t
\leq \Delta x$. In addition, the source term from the derivative of
the potential $\partial V/ \partial \Phi$ adds an effective mass
term to the equations. For stability, the timestep must be short
enough that variations in the field are well resolved. One way to see
this is to look at a vector theory in the situation where the field is
spatially constant. We then have an equation
$$
  \ddot\Phi = -\lambda (|\Phi|^2 - v^2)\Phi\ ,\eqno(3.5)
$$
where the effective mass is $\lambda (|\Phi|^2 - v^2)$. If we
normalize $v$ to be $1$, then when $\Phi$ is close to zero and at
the core of the potential, the effective mass is at a maximum, and the
solution to the above differential equation has a wavenumber of
$\lambda^{1/2}$. To resolve the solution for this wavenumber, we
typically take the timestep to be $1/10$ of the wavelength.

For the parameters of the scalar potential, we take the quartic
self-coupling of the field to be $\lambda=0.1$, and the magnitude
of the vacuum expectation value to be $v=0.1$.  The $\so5/\so4$ model 
contains the additional parameter $a$, which characterizes where on the
vacuum manifold $S^4$ the initial configuration lies; we have
used $a=0.1$ in the simulations shown.  (Additional simulations with
other values of $a$ led to indistinguishable results.)
Together with a timestep $\Delta t \simeq 0.1$ our code is stable for 
all collapses in the theories studied.

To ensure stability and the accuracy of the code, we check that the
total energy within the simulation volume is constant. We find that
for the above parameters, the energy is typically constant to within
$1\%$. There is often a slight increase due to the production of massive
radiation. This is due to an increase in the effective mass of the
field, and subsequently a poorer resolution of fluctuations in the
field (discussed above) in time.

\noindent{\bf IV. Results}

For each of the eight models listed in Table One, we performed
three simulations of single-texture collapse; one with a spherically
symmetric initial configuration, one with an oblate initial configuration,
and one with a prolate initial configuration.  The oblate and prolate
configurations were obtained by scaling the configuration along either
one or two axes by a factor of 0.5.  
At each timestep in the simulations we calculate the total energy $E$,
as well as the separate values of the kinetic energy $E_k$,
gradient energy $E_g$, and potential energy $E_V$:
$$
  \eqalign{ E = & E_k + E_g + E_V\cr
  E_k = & \int d^3x\ \left({{\partial\Phi}\over{\partial t}}\right)^2\cr
  E_g = & \int d^3x\ \left(\nabla\Phi\right)^2 \cr
  E_V = & \int d^3x\ V(\Phi)\ .\cr}\eqno(4.1)
$$
To measure the asymmetries in the collapse, we computed the
quadrupole moments of the total energy in a sphere of radius the
initial texture size:
$$
  Q_{ij} = \int_{r<R} d^3x\ (3x^i x^j - \delta^{ij} r^2)\rho({\bf x}) ,
  \eqno(4.2)
$$
where $\rho$ is the total energy density.  Since we set up our
configurations with initial asymmetries along the coordinate
axes, we are interested in the moments $Q_{ii}$ which measure the
asymmetry along the $x^i$ axis.

We examine first the four models listed in Table One which do
not support strings or monopoles; $\p1(\M) = \p2(\M) = 0$.  
In Figure Two we have plotted the integrated energies (total,
gradient, kinetic and potential) for the evolution of initially
symmetric configurations in each of these theories.
The total energy is normalized to $100$ for purposes of comparison.
These graphs reveal several features which are common to all
of the collapse simulations we studied.  The configuration starts
out with zero kinetic and potential energy.  As the texture
collapses the potential energy
grows slightly, but remains a small part of the total energy
throughout the simulation.  The kinetic and gradient energies
move toward equipartition, although this is not necessarily
completely successful.  In every simulation we performed, 
the graphs of the kinetic and gradient energies approach each
other and cross at least once; the point where they are first equal 
serves as a convenient measure of the timescale for
the collapse.  In the simulations shown in Figure Two, equality
is reached after approximately 75 timesteps.

A remarkable feature of these four graphs is their similarity
with respect to each other.  In the $\so4/\so3$ and $\so5/\so4$
theories the kinetic and gradient energies do not equipartition
quite as effectively as in the $\so5/\so3$ and $\su3/0$ theories,
but the difference is small.  As we shall see below, the differences
within this set of models are much less than the differences between
them and the models which support defects (strings or monopoles).

The results of the $\so5/\so4$ simulation are especially noteworthy,
since in that theory the initial configuration is topologically
trivial [$\p3(\so5/\so4)=0$].  Nevertheless, the evolution of the
energy densities proceeds in a manner practically indistinguishable
from that in $\so4/\so3$.  (The fact that such configurations would
collapse was noted in [2]; our simulations provide dynamical evidence
that the collapse is almost identical to those in theories with nontrivial
$\p3(\M)$.)  The only noticeable difference is in the
potential energy, which remains essentially zero throughout the
$\so5/\so4$ collapse, while rising slightly in the other models.
This is due to the fact that, in the theories with topologically 
nontrivial configurations, the fields must climb out of the
vacuum manifold in order to unwind, which necessarily induces a
nonzero potential energy.  Since the potential energy is nevertheless
only a small contributor to the total energy of the texture, these
simulations indicate that the behavior characteristic of ``texture
collapse'' may arise even without nontrivial topology.  (Of course
the frequency with which such behavior actually does arise in a 
cosmological context is not addressed by our simulations, and may
be different in the different models.)

We have not shown the evolution of the quadrupole moments for 
these simulations, since they vanish throughout the course
of the collapse.  This simply indicates that these models remain
spherically symmetric if they begin in a symmetric configuration.

In Figure Three we have plotted the same quantities as in Figure Two,
in the same set of models,
this time for initially oblate configurations.  (The results for
prolate collapse will not be shown, as they are essentially identical 
to the oblate case.)  Once again, the results of these four
simulations are very similar to one another.  They are also similar
to the results for initially symmetric configurations; again we see
the equipartition of kinetic and gradient energies, while the 
potential energy is much smaller.  The timescale for these collapses
is slightly less than that in the symmetric case; the kinetic and
gradient energies first become equal after approximately 60 timesteps.
This difference can be traced to the fact that the initial configurations 
are somewhat smaller, as they are obtained from the symmetric 
configurations by shrinking by $50\%$ along the $x$-axis.

The quadrupole moments for the evolution of the initially oblate 
configurations in models without strings or monopoles are shown in
Figure Four.  The three quantities $Q_{xx}$, $Q_{yy}$ and $Q_{zz}$
are plotted; only two curves are visible because the initial deformation
is along the $x$-axis, and hence $Q_{yy}=Q_{zz}$ throughout the
simulation.  The quadrupoles begin at a certain value, head toward
zero, and cross the axis before settling smoothly toward zero.
The physical interpretation is that the initially oblate configuration
moves toward spherical symmetry, then overshoots to become somewhat
prolate. The quadrupole moments are only calculated in a sphere of
radius equal to the initial texture size, 
so, as the radiation leaves the
sphere, the quadrupole goes to zero. The remarkable resemblance
between the different simulations is the strongest evidence that
texture collapse is essentially the same in these four models; not
only do the total energies partition in similar ways, but the spatial
distribution of the total energy density for nonspherical initial
configurations also evolves similarly.

We turn next to the four models listed in Table One which support
strings or monopoles.  (The initial configurations that we set up
are pure texture, since the fields are everywhere in the vacuum
manifold, but these models allow topologically
stable defect configurations.)  In Figure Five we have plotted the 
integrated energies (total, gradient, kinetic and potential) for
the collapse of configurations with initially spherically symmetric 
energy densities in
these models.  The four graphs are similar to each other, although
we shall see below that these models are distinguishable by other
measures.  As in the models that do not support defects,
the kinetic and gradient energies evolve toward equipartition, while
the potential energy is always significantly smaller.  A comparison
between Figures Two and Five, however, reveals that there are
important differences between the spherical collapses in models
with and without strings or monopoles.  The most significant
quantifiable difference is in the timescale for collapse; in the models
which allow defects, the kinetic and gradient energies become equal
after only 45 timesteps, in comparison with 75 timesteps for the
models of Figure Two.  

Another dramatic difference is manifested in Figure Six, which
shows the quadrupole moments for the collapses of initially
spherical configurations in models which support defects.  The notable
feature of these plots is that
the quadrupoles in three of the four models [$\so3/\so2$, $\so3/0$, and
$\so4/\u2$] do not vanish throughout the simulation, even though the initial
energy densities are spherically symmetric.  As discussed at the end of 
Section Two, the asymmetries indicated by the nonzero quadrupoles 
can develop because the initial field configuration is
not truly spherically symmetric, even though the energy density is.
The collapse of these textures is therefore different in an important
way from the collapse of truly symmetric textures, since the 
energy densities evolve asymmetrically.  On the other hand, it is
also important to recognize that in the remaining model,
$\so5/[\so3\times\so2\times\Z_2]$, which is truly symmetric and does
not develop a nonzero quadrupole, the partition of energy densities
nevertheless resembles the other models which support defects more
than those which do not.  This is why it is more appropriate to
classify the theories by whether or not they support strings or
monopoles, rather than whether or not they admit spherically 
symmetric texture configurations.

Closer examination of the field values during these simulations
reveals that the asymmetry leads to dramatic effects, in the form of
nucleation of topological defects.  Figure Seven is a three-dimensional
rendering of contours of total energy density and potential energy
at a single timestep during the $\so3/\so2$ simulation.  This theory
supports monopoles as well as textures, and the disk-shaped regions
of high potential energy are monopoles and antimonopoles.  A sequence
of such pictures throughout the simulation reveals that the texture
initially collapses in a nearly symmetric fashion.  It then begins to
deviate from spherical symmetry (while remaining axially symmetric),
and a monopole-antimonopole pair nucleates at the center.  These
travel away from each other toward the edges of the simulation region,
while eventually another pair nucleates at the center.  The antimonopole
of the second pair follows the monopole of the first, and vice-versa.
Presumably these will annihilate, although the first
pair leaves the box before this can occur.  The simulation of 
$\so4/\u2$ proceeds along the same lines, with identical results.
This is not surprising, since the vacuum manifold $\so4/\u2 = {\bf R}P^2$
is related to the vacuum manifold $\so3/\so2 = S^2$ by identification of
antipodal points; the local geometry of both vacuum manifolds is 
therefore the same.  Laboratory experiments with nematic liquid
crystals [11] have seen evidence of similar behavior; the vacuum
manifold in that case was ${\bf R}P^2$, and it was found that textures
tended to decay into monopole-antimonopole pairs.
Meanwhile, the distinct geometry of $\so3/0 =\R P^3$ leads to
dramatically different evolution; in this model there are no
monopoles, but the texture collapse leads to formation of a loop
of cosmic string, as shown in Figure Eight.  This loop grows and
intersects the edges of the simulation region, so we cannot follow
its entire evolution.  The $\so5/[\so3\times\so2\times\Z_2]$ theory,
although admitting both strings and monopoles, collapses symmetrically,
and no defects are formed.  

Of course, in a cosmological context it would be very unlikely for
a collapsing texture to begin in a perfectly symmetric configuration.
To this extent, the instability of some theories 
is of less importance than the evolution of initial
conditions which are noticeably asymmetric.  Figure Nine presents the
behavior of the integrated energies in the four models which support
defects, beginning from oblate initial conditions.  Again, the
four plots appear similar to each other, while differing from the
analogous plots in Figure Three for models without strings or monopoles.
The timescale for collapse is now approximately 35 timesteps, and
the equipartition of kinetic and gradient energies occurs rapidly
and effectively.

Figure Ten, which plots the quadrupole moments for these collapses
from oblate initial conditions, indicates that the models with
defects can differ from each other as well as from the models without
defects.  (Although it is not evident from these plots, we have
checked that defects are nucleated according to the same pattern
which appeared in symmetric collapses; monopole/antimonopole
pairs are produced in $\so3/\so2$ and $\so4/\u2$, a string loop is
produced in $\so3/0$, and no defects are produced in 
$\so5/[\so3\times\so2\times\Z_2]$.)  The general pattern familiar
from Figure Four, where the oblate configuration became prolate
before symmetrizing, is seen here as well.  The quadrupole moments 
in the $\so3/\so2$ and $\so4/\u2$
models, whose vacuum manifolds are locally equivalent, are
indistinguishable, but these two appear different from $\so3/0$
and $\so5/[\so3\times\so2\times\Z_2]$, which also appear slightly
different from each other.  As these differences represent 
distinct evolutions of the spatial distribution of the total
energy density, they indicate the possibility that the different
models could lead to distinguishable cosmological density fluctuations.

\noindent{\bf V. Discussion}

We have examined the dynamics of texture collapse in a variety
of scalar field theories with spontaneously broken global symmetries.
Our goal was to determine whether the geometry and topology of
the vacuum manifold would affect the evolution of analogous field
configurations in different models.  The answer is clearly ``yes'';
however, the cosmological significance of the effects observed is
still unclear.

We have been led to two basic results.  First, the collapse of a single 
texture configuration depends on whether the theory under consideration
admits the existence of cosmic strings or monopoles.  Theories
which do allow such defects have textures which collapse somewhat
more rapidly, and in the process of collapse may nucleate loops
of string or monopole/anti\-monopole pairs.  Second, some theories 
with nontrivial $\p3(\M)$ admit textures with spherically symmetric
energy densities, but not texture configurations which are strictly
spherically symmetric.  Such configurations collapse nonspherically.
These two results are related; it is the configurations without true
spherical symmetry that nucleate defects during collapse.  In
the wide class of models considered in [5], all of
the theories without monopoles or strings either allow spherically
symmetric textures or do not have textures at all.

{}From the point of view of cosmology, texture collapse is not the
only important process in a theory which admits strings or monopoles;
in such models the Kibble mechanism predicts that there will be an
appreciable number of defects per Hubble volume, and the gravitational
perturbations produced by these defects will presumably be comparable
in importance to those produced by collapsing textures.  Indeed, the
results of our simulations may indicate that textures are even less
important in such models than might have been naively expected.
The laboratory experiments of [11], involving a condensed-matter
system which admitted strings, monopoles and textures, found that
textures were rarely produced in a phase transition, and when a 
texture configuration was introduced by hand it decayed into 
monopole/antimonopole pairs.  Our results, in which textures with
initially perfectly symmetric energy densities were seen to collapse
into defects, may indicate that textures will rarely form in certain
theories, with defect formation being more likely.
Contrariwise, the collapse of
textures may provide an additional mechanism for defect production.
The importance of these effects to cosmological structure formation
can only be answered by large-scale numerical simulations of the
evolution of the fields as the universe expands.

For the models which are free of strings and monopoles, our results
provide evidence that simulations of any specific simple model
[such as $\so4/\so3$] do not miss any important physical effects.
We find that the collapse of textures proceeds in essentially the
same way in all such theories; to this extent, simulations in one
model are relevant to other models as well.  Of course, there are
a number of important issues which our simulations do not address,
such as the frequency and distribution of texture collapse, which
could in principle be different in distinct models.

Finally, our examination of the $\so5/\so4$ model provides
evidence that events resembling texture collapse can occur in the
absence of nontrivial topology, as suggested in [2].  Indeed, the
quantitative characteristics of the collapse in this model are
indistinguishable from those in the conventional $\so4/\so3$ model.
Again, the relative frequency of such events in the different 
models is not addressed by our simulations.  More detailed studies
will be necessary to understand the precise differences between
cosmological models based on the theories examined here.

\noindent{\bf Acknowledgments}

We would like to thank Peter Arnold, Jim Bryan,
Edward Farhi, Gary Gibbons, Nick 
Manton, Krishna Rajagopal, Mark Trodden, Neil Turok, Alex Vilenkin,
and Lawrence Yaffe for helpful conversations.  This work
was supported in part by the National Science Foundation under grants
PHY/94-07194 and PHY/92-00687, by the U.S. Department
of Energy (D.O.E.) under contract no. DE-AC02-76ER03069, and by
U.K. PPARC grant GR/L21488.

\vfill\eject

\noindent{\bf References}

\item{1.} N. Turok, {\it Phys. Rev. Lett.} {\bf 63}, 2625 (1989).

\item{2.} L. Perivolaropoulos, {\it Phys. Rev. D} {\bf 46}, 1858
(1992).

\item{3.} U. Pen, D.N. Spergel, and N. Turok, {\it Phys. Rev. D}
{\bf 49}, 692 (1994). 

\item{4.} L.-F. Li, {\it Phys. Rev D} {\bf 9}, 1723 (1974).

\item{5.} J.A. Bryan, S.M. Carroll and T. Pyne, {\it Phys. Rev. D}
{\bf 50}, 2806 (1994), hep-ph/9312254.

\item{6.} D.N. Spergel, N. Turok, W.H. Press, and B.S. Ryden,  
{\it Phys. Rev. D} {\bf 43}, 1038 (1991);  R.Y. Cen, J.P.
Ostriker, D.N. Spergel, and N. Turok, {\sl Ap. J} {\bf 393},
42 (1992); 
D.P. Bennett and S.H. Rhie, {\it Astrophys. J.} {\bf 406}, L7 (1993), 
hep-ph/9207244; R. Durrer, A. Howard, and Z.-H. Zhou, {\it Phys. Rev. D}
{\bf 49}, 681 (1994), astro-ph/9311040;
J. Borrill, E.J. Copeland, A.R. Liddle, A. Stebbins, and S.Veeraraghavan,
{\it Phys. Rev. D} {\bf 50}, 2469 (1994), astro-ph/9403005;
R. Durrer and Z.-H. Zhou, {\it Phys. Rev. Lett.} {\bf 74}, 1701 (1995),
astro-ph/9407027; N.G. Phillips and A. Kogut, {\it Phys. Rev. Lett.} 
{\bf 75}, 1264 (1995), astro-ph/9507045.

\item{7.} J. Borrill, E.J. Copeland, and A.R. Liddle, {\it Phys.
Lett.} {\bf B258}, 310 (1991); R.A. Leese and T. Prokopec, 
{\it Phys. Rev. D} {\bf 44}, 3749 (1991);
S. Aminneborg, {\it Nucl. Phys.} {\bf B388}, 521 (1992);
A. Sornborger, {\it Phys. Rev. D} {\bf 48}, 3517 (1993),
astro-ph/9303005.

\item{8.} T. Prokopec, A. Sornborger,
and R.H. Brandenberger, {\it Phys. Rev. D} {\bf 45}, 1971 (1992).

\item{9.} M. Joyce and N. Turok, {\it Nucl. Phys.} {\bf B416},
389 (1994), hep-ph/9301287.

\item{10.} S.H. Rhie and D.P. Bennett, preprint UCRL-JC-110560 (1992);
X. Luo, {\it Phys. Lett.} {\bf B287}, 319 (1992).

\item{11.} I. Chuang, R. Durrer, N. Turok, and B. Yurke,
{\it Science} {\bf 251}, 1336 (1991).

\vfill\eject

\settabs8\columns

\vskip .5cm
\hrule
\vskip .1cm
\hrule
\vskip .2cm
\+ $G$ & $H$ && $N$ & $\p1(G/H)$ & $\p2(G/H)$ & $\p3(G/H)$ & dim($G/H$) \cr
\vskip .2cm
\hrule
\vskip .2cm
\+ $\so{4}$ & $\so3$ && 4 & 0 & 0 & \Z & $3$ \cr
\+ $\so{5}$ & $\so{4}$ && 5 & 0 & 0 & 0 & $4$ \cr
\+ $\so{5}$ & $\so3$ && 10 & 0 & 0 & \Z$_2$ & $7$ \cr
\+ $\su{3}$ & 0 && 12 & 0 & 0 & \Z & $8$ \cr
\vskip .1cm \hrule \vskip .1cm
\+ $\so{3}$ & $\so2$ && 3 & 0 & \Z & \Z & $2$ \cr
\+ $\so{3}$ & 0 && 6 & \Z$_2$ & 0 & \Z & $3$ \cr
\+ $\so4$ & $\u2$ && 6 & $\Z_2$ & \Z & $\Z$ & 2 \cr
\+ $\so{5}$ & $\so{3}\times\so{2}\times\Z_2$ &&
    14 & $\Z_2$ & $\Z$ & $\Z_2$ & $6$ \cr

\vskip .2cm
\hrule
\vskip .1cm
\hrule
\vskip .4cm

\noindent{\bf Table 1: Models.}  We list the models considered
in this paper, in which a global symmetry group $G$ is spontaneously
broken to a subgroup $H$.  $N$ is the number of real scalar fields;
the twelve real fields in the $\su3$ theory are arranged into two
complex three-vectors, while the other representations are explicitly
real.  The lower homotopy groups $\p{q}$ of the vacuum manifold
$G/H$ are listed, as well as the dimensionality of $G/H$ (corresponding
to the number of massless Goldstone bosons).  The models are grouped
into two classes, depending on whether the model supports strings
[$\p1(G/H)$] and/or monopoles [$\p2(G/H)$].

\vfill\eject

\centerline{\bf Figure Captions}

\noindent{\bf Figure One.}  These plots represent energy density 
contours in the $x$-$y$ plane of oblate configurations in
the $\so4/\so3$ and $\so3/\so2$ theories.  The textures are
obtained from those with spherically symmetric energy densities by
shrinking the configurations by $50\%$ along the $x$-axis.
The energy density in the $\so4/\so3$ is simply deformed along 
with the coordinates, while in the $\so3/\so2$ model the maximum
energy density describes a ring in the $y$-$z$ plane.  In these
figures, the three-dimensional configuration is obtained by 
rotation around the $x$-axis.

\noindent{\bf Figure Two.}  The integrated energy densities during
the evolution of textures with initially spherically symmetric energy
densities in the four theories which do not support strings or 
monopoles.  Each form of energy density is integrated over the volume
of the simulation region; the total energy is the sum of the integrated
potential, kinetic and gradient energies.  The horizontal axis is
measured in timesteps of the simulations.

\noindent{\bf Figure Three.}  The integrated energy densities during
the evolution of textures with initially oblate configurations (obtained
from the symmetric configurations by deforming along one axis)
in the four theories which do not support strings or monopoles. 

\noindent{\bf Figure Four.}  Quadrupole moments of the total energy
density for collapse of initially oblate configurations
in the four theories which do not support strings or monopoles. 

\noindent{\bf Figure Five.}  The integrated energy densities during
the evolution of textures with initially spherically symmetric energy
densities in the four theories which do support either strings or 
monopoles.  (These textures are referred to as ``symmetric'',
although only in $\so5/[\so3\times\so2\times\Z_2]$ is the configuration
truly spherically symmetric.)

\noindent{\bf Figure Six.}  Quadrupole moments of the total energy
density for collapse of configurations with initially symmetric 
energy densities in the four theories which do support either strings or 
monopoles.  Note that nonzero quadrupoles develop for three of the four
models.

\noindent{\bf Figure Seven.}  Three-dimensional contours of total
energy density and potential energy at one moment during the
collapse of the $\so3/\so2$ configuration with an initially symmetric energy
density.  The shaded regions are those with large potential energies,
representing monopoles and anti-monopoles.  The lower total-energy
contour describes a prolate spheroid, while the higher contour
describes two roughly hemispherical regions associated with the inner
monopole/anti-monopole pair.

\noindent{\bf Figure Eight.}  Three-dimensional contours of total
energy density and potential energy at one moment during the
collapse of the $\so3/0$ configuration with an initially symmetric energy
density.  The shaded region is one of large potential energy,
representing a loop of cosmic string.  The total-energy contours
describe roughly spheroidal regions.

\noindent{\bf Figure Nine.}  The integrated energy densities during
the evolution of textures with initially oblate energy densities (obtained
from the symmetric configurations by deforming along one axis)
in the four theories which do support either strings or monopoles.  

\noindent{\bf Figure Ten.}  Quadrupole moments of the total energy
density for collapse of initially oblate configurations
in the four theories which do support either strings or monopoles.

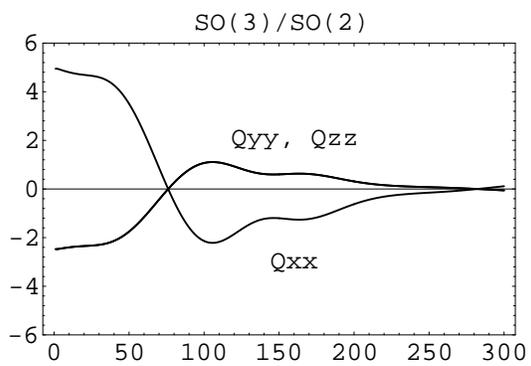
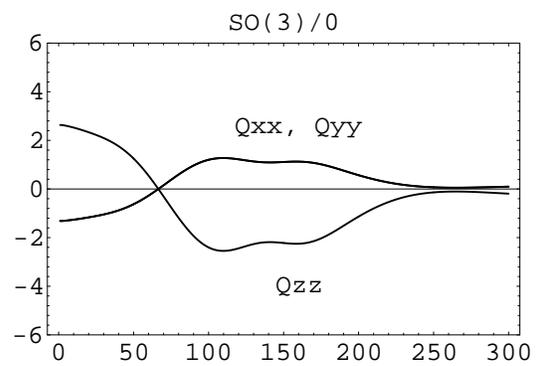
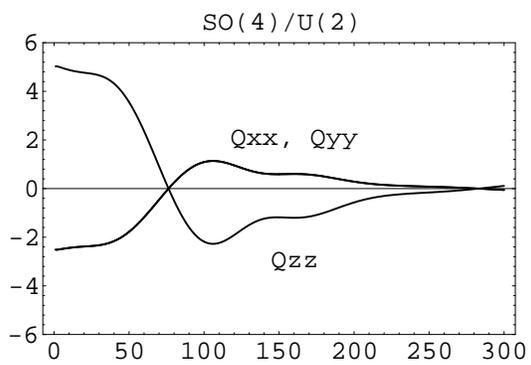
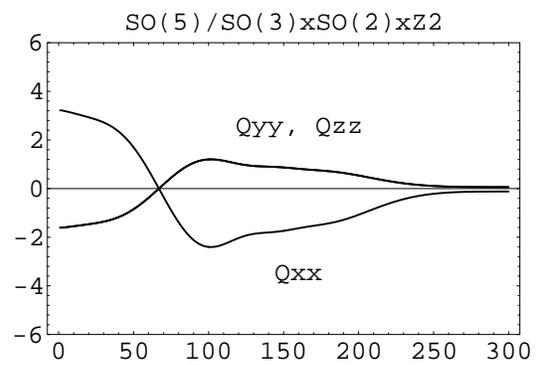
\bye